\documentclass[aps,prl,twocolumn,showpacs,superscriptaddress,amssymb,amsmath]{revtex4-1} 
\usepackage{latexsym}
\usepackage[dvips]{graphicx}
\usepackage{bm}
\usepackage{epsfig}
\usepackage{dcolumn}

\begin{document}

\title{X-ray cross-correlation analysis of liquid crystal membranes in the vicinity of hexatic-smectic phase transition}

\author{R.P. Kurta}
\affiliation{Deutsches Elektronen-Synchrotron DESY, Notkestra\ss e 85, D-22607 Hamburg, Germany}
\author{B.I. Ostrovskii}
\affiliation{Institute of Crystallography, Russian Academy of Sciences, 119333 Moscow, Russia}
\author{A. Singer}
\altaffiliation[New address: ]{The University of California, San Diego, La Jolla, CA 92093, USA}
\affiliation{Deutsches Elektronen-Synchrotron DESY, Notkestra\ss e 85, D-22607 Hamburg, Germany}
\author{O.Y. Gorobtsov}
\affiliation{Deutsches Elektronen-Synchrotron DESY, Notkestra\ss e 85, D-22607 Hamburg, Germany}
\author{A. Shabalin}
\affiliation{Deutsches Elektronen-Synchrotron DESY, Notkestra\ss e 85, D-22607 Hamburg, Germany}
\author{D. Dzhigaev}
\affiliation{Deutsches Elektronen-Synchrotron DESY, Notkestra\ss e 85, D-22607 Hamburg, Germany}
\author{O.M. Yefanov}
\affiliation{Center for Free-Electron Laser Science CFEL, Notkestra\ss e 85, D-22607 Hamburg, Germany}
\author{A.V. Zozulya}
\affiliation{Deutsches Elektronen-Synchrotron DESY, Notkestra\ss e 85, D-22607 Hamburg, Germany}
\author{M. Sprung}
\affiliation{Deutsches Elektronen-Synchrotron DESY, Notkestra\ss e 85, D-22607 Hamburg, Germany}
\author{I.A. Vartanyants}
\email[Reference author: ]{ivan.vartaniants@desy.de}
\affiliation{Deutsches Elektronen-Synchrotron DESY, Notkestra\ss e 85, D-22607 Hamburg, Germany}
\affiliation{National Research Nuclear University, ``MEPhI'', 115409 Moscow, Russia}
\date{\today}

\begin{abstract}
We present an x-ray study of liquid crystal membranes in the vicinity of hexatic-smectic phase transition by means of
angular x-ray cross-correlation analysis (XCCA). By applying two-point angular intensity cross-correlation functions to the measured series of
diffraction patterns the parameters of bond-orientational (BO) order in hexatic phase were directly determined.
The temperature dependence of the positional correlation lengths was analyzed as well.
The obtained correlation lengths show larger values for the higher-order Fourier components of BO order.
These findings indicate a strong coupling between BO and positional order that has not been studied in detail up to now.

\end{abstract}


\maketitle

\section{Introduction}

About 30 years ago it was realized that the phase transition between a two-dimensional crystal and liquid phase can proceed through an intermediate hexatic phase \cite{Halperin}.  The corresponding mechanism involves dissociation of dislocation pairs. The two-dimensional (2D) hexatic phase is characterized by a six-fold quasi-long-range bond-orientational (BO) order, while the positional order is short range and the shear modulus is zero. Phases with hexatic order have been found in several systems, such as electrons at the surface of helium \cite{Glattli}, charged polymer colloids \cite{Murray, Kusner} and smectic liquid crystals (LCs) \cite{Strandburg, Brock1, Stoebe1}. In the later case a hexatic phase was experimentally observed quite unexpectedly in 3D stacked molecular systems \cite{Pindak}. In 3D hexatic phase the positional order is short range, while the BO order persists over long-range. However, the mere existence of a hexatic phase does not imply a specific melting mechanism, and the origin of hexatic order and its relation to defect-mediated melting transition is still controversial. Smectic liquid crystals are particularly suitable to investigate these problems, as they can be suspended over an opening in a solid frame. Such smectic membranes are substrate free and have a controlled thickness ranging from two to over thousands of layers \cite{Wim}.

Smectic A membranes can be described as stacks of liquid layers. The in-plane structure is liquid-like with positional correlations between the molecules decaying exponentially with a correlation length $\xi_{0}$. While cooling a hexatic phase may occur, which shows a long-range BO order. This leads to a six-fold rotational symmetry and the BO correlations are characterized by the local ordering field  $\psi(\mathbf{r})\propto\exp[i6\theta(\mathbf{r})]$, where $\theta(\mathbf{r})$ is the angle between the ``bonds'' and some reference axis. Upon decreasing temperature, the width of the radial intensity peak decreases simultaneously with a further development of the BO
order.
This indicates a coupling between the positional correlations and the BO order \cite{Aeppli}.
Additionally, at even lower temperatures a 3D crystalline phase appears with a hexagonal in-plane lattice or a rectangular lattice with a so-called herringbone order \cite{Brock1,Stoebe1}.

The common way to study the BO order is to perform x-ray or electron diffraction measurements on a single-domain hexatic film at different temperatures \cite{Chou, Brock}.  Information about the temperature evolution of the BO order parameters is typically obtained by fitting the measured azimuthal intensity distribution. We propose a different approach based on angular x-ray cross-correlation analysis (XCCA) \cite{Wochner1, Altarelli,*[{Erratum: }]AltarelliErr, Kurta1, Kurta2, Kurta3} that allows to determine the BO order parameters directly, without fitting. Angular XCCA has been developed to study the structure of non-crystalline systems like liquids, amorphous systems, disordered ensembles of particles, etc. Particularly, it can be used to identify local structures or to detect hidden symmetries in disordered systems from the study of angular correlations of the diffracted intensity.
Smectic membranes are especially suited for Fourier analysis of intensity cross-correlation functions (CCFs). They are not influenced by substrate interactions and their two surfaces induce an almost perfect 2D alignment of the smectic layers.

In this paper we aim to study a continuous crossover from disordered smectic phase to hexatics, showing 6-fold BO order. We perform XCCA of series of diffraction patterns measured at different spatial positions of the film at each temperature. Applying Fourier analysis of the averaged two-point CCFs we directly determine the BO order parameters and their temperature evolution, as well as corresponding positional correlation lengths in the vicinity of smectic-hexatic phase transition.


\section{Theory}

The in-plane structure factor in smectic phase has the form of a broad ring due to short-range positional correlations between molecules.
The presence of BO order in hexatic phase breaks the angular isotropy of the structure factor and leads to a six-fold modulation of the in-plane scattering. The angular Fourier decomposition
of the intensity $I(q,\varphi)$ scattered at the momentum transfer vector $\mathbf{q}$ can be defined as
\begin{equation}
I(q,\varphi)=I_{0}(q)+2\sum\limits_{n=1}^{\infty}I_{n}(q)\cos(n\varphi),\label{Eq:Cchi}
\end{equation}
where the radius $q=\lvert\mathbf{q}\rvert$ and azimuthal angle $\varphi$ are polar coordinates of $\mathbf{q}$, $I_{n}(q)$ are the Fourier components of intensity $I(q,\varphi)$, with $I_{0}(q)$ representing the angular averaged intensity.
In the case of a single hexatic domain, only the components $I_{n}(q)$ with $n=6m,\;m=1,2,3,4,\dots$ contribute to the expansion (\ref{Eq:Cchi}), while all other components have vanishing values \cite{Brock}.
The values of these Fourier components at $q=q_{0}$, corresponding to the maximum values of $I_{n}(q)$ as a function of $q$, are related to the BO order parameters \footnote{The following decomposition of the azimuthal intensity to describe BO order in LCs was used previously \cite{Brock, Brock1}, $I(\varphi)=I_{0}[1/2+\sum\limits_{m=1}^{\infty}C_{6m}\cos(6m\varphi)]$.
Here, $I_{0}$ and the BO order parameters $C_{6m}$ are related to the Fourier components $I_{n}(q)$ defined in Eq.~(\ref{Eq:Cchi}) as $I_{0}=2I_{0}(q_{0})$ and $C_{6m}=I_{n}(q_{0})/I_{0}(q_{0}),\;n=6m$, with $q_{0}$ corresponding to the maximum of $I_{n}(q)$.} defined in Refs.\cite{ Brock1,Brock}.

\begin{figure*}[!htbp]
\centering
\includegraphics[width=0.6\textwidth]{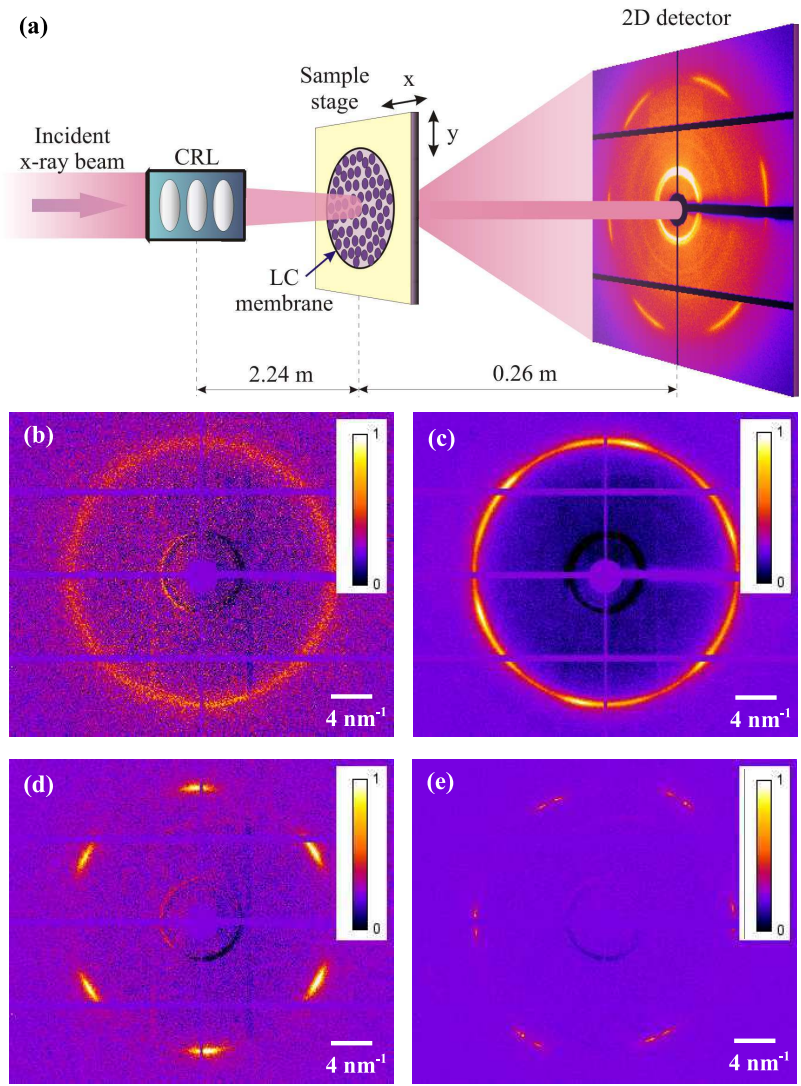} 
\caption{
(a) Geometry of the scattering experiment showing the CRL transfocator, sample stage and 2D detector (b)-(e) Typical diffraction patterns measured for LC membrane at different temperatures $T$.
(b) Smectic phase ($T=64.25^{\circ}\rm{C}$) with a scattering ring at $q_{0}\sim14\;\rm{nm}^{-1}$.
(c) Hexatic phase ($T=62.5^{\circ}\rm{C}$) corresponding to scattering from few domains in different orientations.
(d) Hexatic phase ($T=62.25^{\circ}\rm{C}$) with a prominent six-fold symmetry typical for a single domain.
(e) Crystalline phase ( $T=58.75^{\circ}\rm{C}$) with two domains of slightly different orientation.  Horizontal and vertical stripes on the diffraction patterns are the gaps between the tiles of the Pilatus detector, that were eliminated from the data analysis.
\label{Fig:DiffPat}
}
\end{figure*}

The angular XCCA enables direct determination of the Fourier components $I_{n}(q)$ from the ensemble
of diffraction patterns measured at different positions on the sample \cite{Altarelli,*[{Erratum: }]AltarelliErr, Kurta1}.
For each diffraction pattern a two-point cross-correlation function (CCF) of the form
\begin{equation}
C(q,\Delta) = \left\langle I(q,\varphi)I(q,\varphi+\Delta)\right\rangle _{\varphi} \label{Eq:CCF1}
\end{equation}
is defined, where $\Delta$ is the angular coordinate, and $\langle \dots \rangle_{\varphi}$
denotes the angular average around the ring of a radius $q$.
The CCFs $C(q,\Delta)$ can be analyzed using a Fourier series decomposition \cite{Altarelli,Kurta1},
\begin{equation}
C(q,\Delta)=C_{0}(q)+2\sum\limits_{n=1}^{\infty} C_{n}(q)\cos(n\Delta),\label{Eq:Cqn1}
\end{equation}
where $C_{n}(q)$ are the Fourier components of the CCF, particularly $C_{0}(q)\equiv \langle I(q,\varphi)\rangle_{\varphi}^{2}$.

In the case of scattering from several domains, the values of the Fourier components $C_{n}(q)$ may fluctuate from position to position on the membrane \cite{Altarelli,*[{Erratum: }]AltarelliErr, Kurta1, Kurta2}.
However, as demonstrated in the previous work \cite{Kurta1, Kurta2}, averaging the Fourier components of the CCF $\langle C_{n}(q) \rangle_{M}$ over a sufficient number $M$ of diffraction patterns leads
to the explicit relation \footnote{The interference terms from different domains were neglected in derivation of Eq.~(\ref{Eq:CqnAver1}).} between $\langle C_{n}(q) \rangle_{M}$ and $I_{n}(q)$,
\begin{equation}
\langle C_{n}(q) \rangle_{M}=K\lvert I_{n}(q)\rvert^2,
\label{Eq:CqnAver1}
\end{equation}
where $I_{n}(q)$ are the Fourier components of the angular expansion of intensity $I(q,\varphi)$ corresponding to a single domain, and $K$ is a scaling coefficient that depends on the number of domains
contributing to diffraction patterns and distribution of their orientations \cite{Kurta1, Kurta2, Kurta3}.
Equation (\ref{Eq:CqnAver1}) can be directly used to determine the magnitudes of the Fourier components $\lvert I_{n}(q)\rvert$.
For a single domain case the scaling coefficient $K=1$ and $\lvert I_{n}(q)\rvert=\langle C_{n}(q)\rangle_{M}^{1/2}$.

\section{Experiment}

The coherent x-ray scattering experiment was performed at the coherence beamline P10 of the PETRA III facility at DESY in Hamburg.
The scattering geometry of the experiment is shown in Fig.~\ref{Fig:DiffPat}(a).
The incident photon energy was chosen to be $13\;\rm{keV}$ and a 2D detector was positioned in transmission geometry at $263\;\rm{mm}$ distance from the sample and protected by a beamstop of $15\;\rm{mm}$ in diameter.
The scattering data were recorded on a hybrid-pixel detector Pilatus 1M from Dectris with $981\times1043$ pixels and a pixel size of $172\times 172\;\mu\rm{m}^{2}$.
A specially designed sample stage FS1 together with mK1000 temperature controller from INSTEC has been used for preparation of LC thin films. Exit apertures of the stage were covered by Kapton foil of $25\;\mu\rm{m}$ in thickness to preserve temperature uniformity inside the chamber.

The smectic membranes of LC compound $n$-heptyl-4'-$n$-pentyloxybiphenyl-4-carboxylate (75OBC)
of different thickness were drawn across a small circular glass aperture of $2\;\rm{mm}$ in diameter inside the chamber at $10-12^{\circ}\rm{C}$ above the temperature of smectic-hexatic phase transition. The thickness of the films was measured using AVANTES optical reflectometry setup, and was in the range $5-8\;\mu\rm{m}$ for different films.
After stabilizing a film at elevated temperature $T$, it was gradually decreased with a temperature ramp $\rm{d}T=0.3^{\circ}\rm{C}/\rm{min}$ to observe a sequence of LC phases [see Fig.~\ref{Fig:DiffPat}(b-e)].

The sample chamber was mounted on a goniometer, and a film was aligned with its surface perpendicular to the direction of the incident beam.
The beam with a flux of about $2\cdot 10^{11}\;\rm{photons/sec}$ was focused on the sample using a beryllium compound refractive lenses (CRLs) \cite{Zozulya} to a spot of about $3\;\mu\rm{m}$ (FWHM).
At each temperature the sample was scanned in the plane perpendicular to the incident beam direction with a step size larger than the probe size.
The exposure times were chosen in the range from $0.2\;\rm{s}$ up to $0.3\;\rm{s}$ per image to perform measurement in non-destructive regime depending on the film thickness.

\begin{figure*}[!htbp]
\centering
\includegraphics[width=0.45\textwidth]{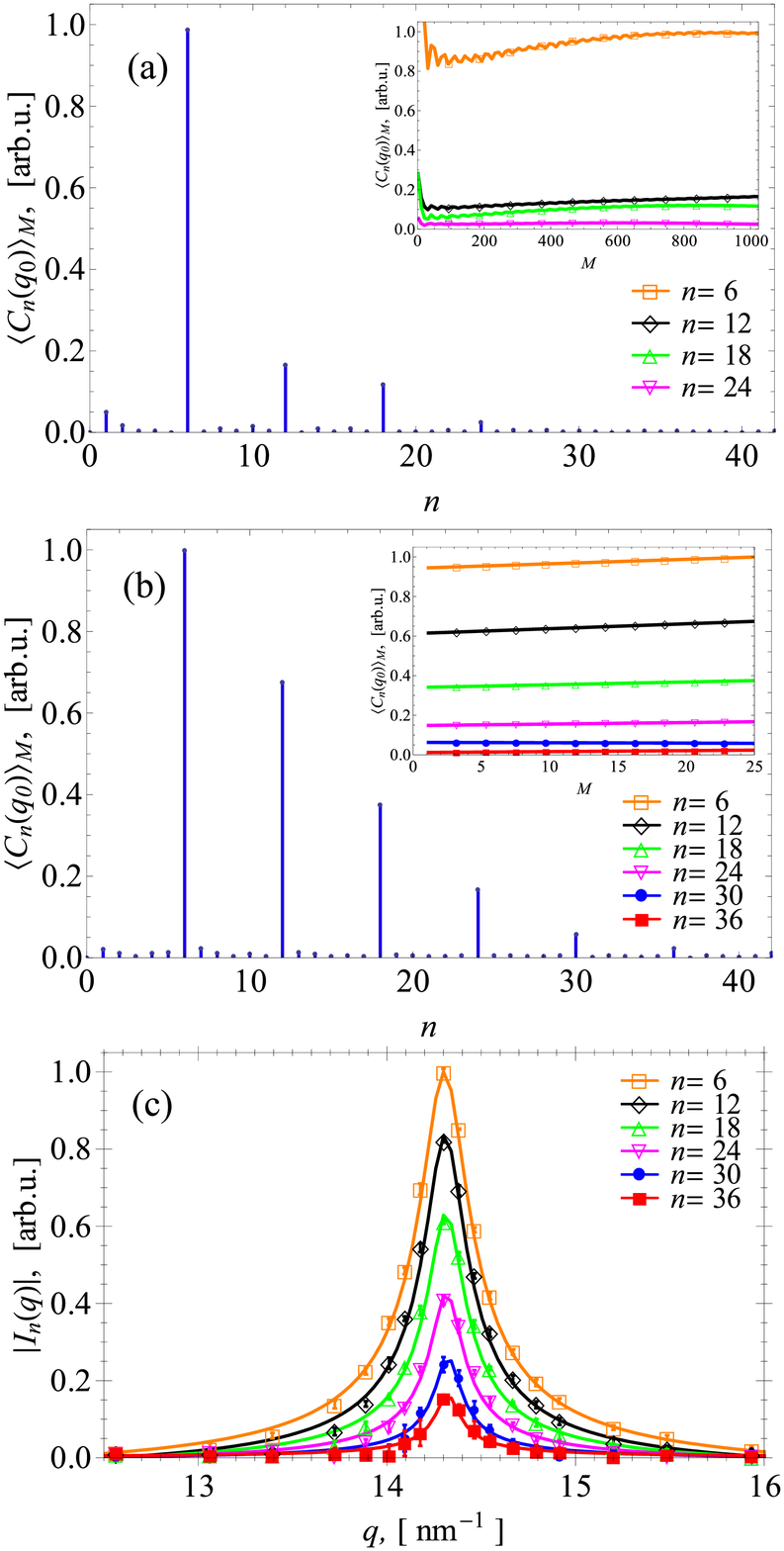} 
\caption{(a),(b) Normalized averaged Fourier components of CCFs $\langle C_{n}(q_{0})\rangle_{M}$ determined at $q_{0}=14.3\;\rm{nm}^{-1}$ for $1 \leq n\leq 40$ for two different smectic membranes, corresponding to a several domain case (a), and a single domain case (b). The insets show evolution of the dominant Fourier components as a function of $M$. (c) Normalized Fourier components $\lvert I_{n}(q)\rvert$ as a function of $q$ determined for the single domain case. Solid lines are SRL fits to the experimental data (points). The error bars for the chosen experimental data points in (c) are obtained by statistical analysis of five individual subensembles ($5$ diffraction patterns each) from the whole set of $M=25$ patterns.
\label{Fig:XCCA1}
}
\end{figure*}
\begin{figure*}[!htbp]
\centering
\includegraphics[width=0.49\textwidth]{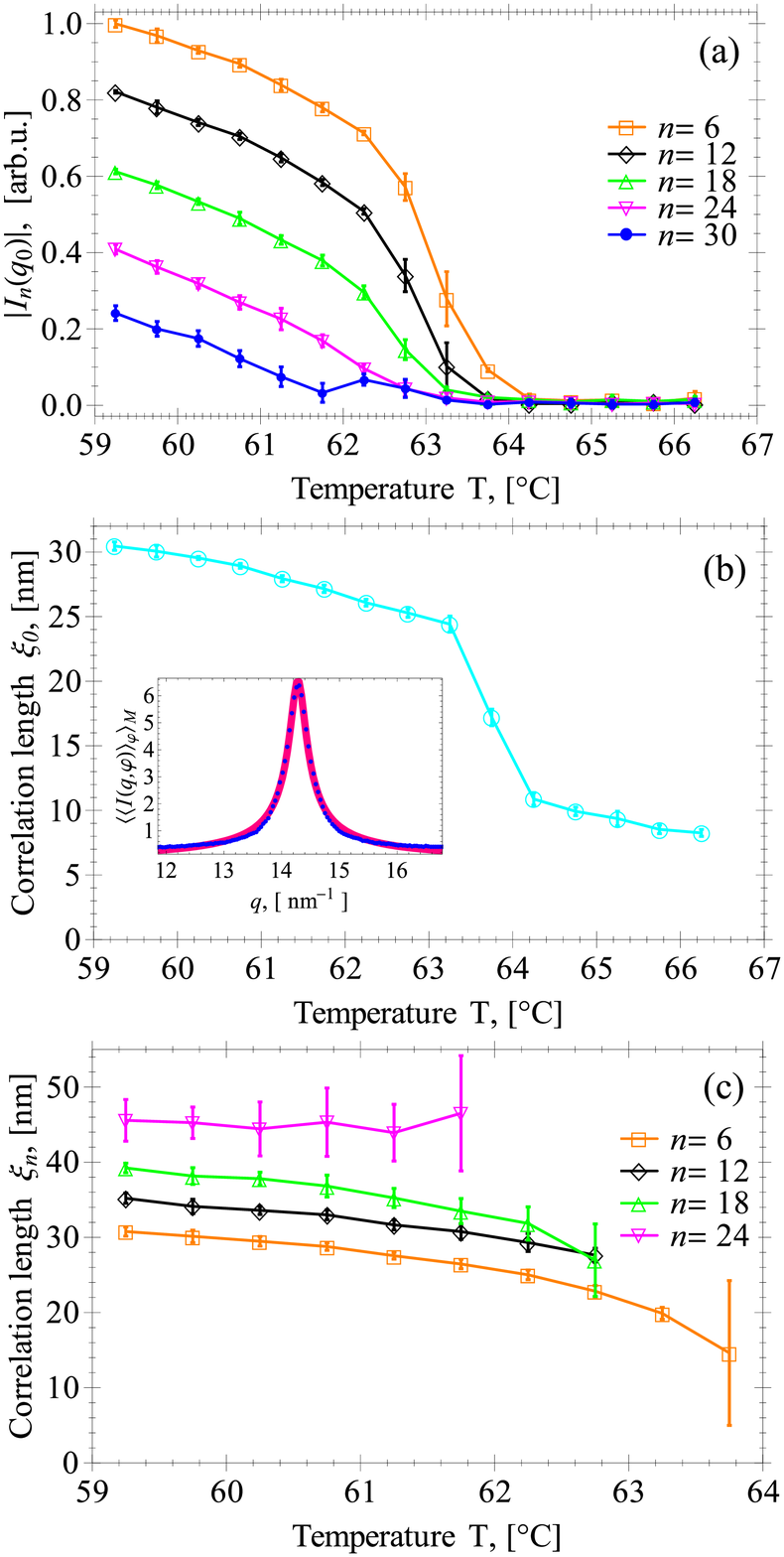} 
\caption{(a) Temperature dependence of BO order parameters $\lvert I_{n}(q_{0})\rvert$ calculated for a single domain case at $q_{0}=14.3\;\rm{nm}^{-1}$.
(b) Temperature dependence of the correlation length $\xi_{0}$, determined from the SRL fits of $\lvert I_{0}(q)\rvert$.
A typical SRL fit for $\lvert I_{0}(q)\rvert\equiv \langle \langle I(q,\varphi)\rangle_{\varphi}\rangle_{M}$ is shown in the inset for $T=63.25^{\circ}\rm{C}$.
(c) Temperature dependence of the correlation length $\xi_{n}$ for $n=6,12,18$ and $24$ obtained similar to (b). Error bars are evaluated in the same way as in Fig.~\ref{Fig:XCCA1}(c).
\label{Fig:XCCA2}
}
\end{figure*}
%


\section{Results and discussion}

For each LC membrane a series of diffraction patterns were recorded at different temperatures.
Measured diffraction patterns were corrected for background scattering and horizontal polarization of synchrotron radiation.
Typical diffraction patterns corresponding to different phases appearing at distinct temperatures are shown in Fig.~\ref{Fig:DiffPat}(b)-\ref{Fig:DiffPat}(e).
Gradually decreasing the temperature from $T=66^{\circ}\rm{C}$ to $58^{\circ}\rm{C}$
intensity distributions typical for smectic [Fig.~\ref{Fig:DiffPat}(b)], hexatic  [Figs.~\ref{Fig:DiffPat}(c),(d)] and crystalline [Fig.~\ref{Fig:DiffPat}(e)] phases were measured. During different transverse scans in $x-y$ plane [see Fig.~\ref{Fig:DiffPat}(a)] we observed cases when the x-ray beam was scattered from a single domain [Fig.~\ref{Fig:DiffPat}(d)], or from several domains [Figs.~\ref{Fig:DiffPat}(c),(e)]. Positions of the Bragg peaks in Fig.~\ref{Fig:DiffPat}(e) correspond to a low-temperature herringbone crystalline E phase with two slightly disoriented domains \cite{Geer1}.

In Fig.~\ref{Fig:XCCA1} the results of correlation analysis [Eqs.~(\ref{Eq:CCF1})-(\ref{Eq:CqnAver1})] are presented for two different LC films.
A large data set with $M=1024$ patterns was measured at the temperature $T=61.5^{\circ}\rm{C}$ for the first film, and a small data set with $M=25$ diffraction patterns at $T=59.75^{\circ}\rm{C}$ for the second film.
Scattering from several domains is indicative for diffraction patterns for the first film, and from a single domain for the second film.
The normalized averaged Fourier components $\langle C_{n}(q)\rangle_{M}$ calculated at $q_{0}=14.3\;\rm{nm}^{-1}$ for $1 \leq n\leq 40$ are presented in Fig.~\ref{Fig:XCCA1}(a) and Fig.~\ref{Fig:XCCA1}(b) for the several and single domain case, respectively. One can readily see, that the dominant contribution to the spectrum is given by the Fourier components with $n=6,12,18,24$, and $n=30,36$ in the single domain case, whereas other Fourier components have vanishing values (except of $n=1$ that we attribute to a small misalignment of the detector). Evolution of the dominant Fourier components $\langle C_{n}(q)\rangle_{M}$ as a function of $M$ is shown in the insets of Figs.~\ref{Fig:XCCA1}(a) and \ref{Fig:XCCA1}(b), demonstrating their statistical convergence.
The normalized Fourier components determined for the single domain case as $\lvert I_{n}(q)\rvert=\langle C_{n}(q)\rangle_{M}^{1/2}$ are presented as a function of $q$ in Fig.~\ref{Fig:XCCA1}(c).
We point out, the $q$-dependence of the Fourier components shown in Fig.~\ref{Fig:XCCA1}(c) is a direct result of application of the XCCA, that makes it advantageous as compared to fitting techniques.

We performed similar analysis at different temperatures and determined the temperature dependence of the BO order parameters
$\lvert I_{n}(q_{0})\rvert$ for a single-domain case. The results of calculations for $q_{0}=14.3\;\rm{nm}^{-1}$, corresponding
to the maximum value of the peaks in Fig.~\ref{Fig:XCCA1}(c), are presented in Fig.~\ref{Fig:XCCA2}(a).
It is readily seen that the hexatic order parameters decay as a power law upon approaching hexatic-smectic transition.
According to the multicritical scaling theory \cite{Aharony}, in the vicinity of the hexatic-smectic transition the
BO order parameters measured for a single domain should follow a power law dependence as a function of $n$.
The results shown in Fig.~\ref{Fig:XCCA2}(a) are in good agreement with this theory
\footnote{The multicritical scaling theory \cite{Aharony}, predicts a power law decay of the BO order parameters $C_{6m}=I_{n}(q_{0})/I_{0}(q_{0}),\;n=6m$ for a single domain system. According to this theory,
$C_{6m} = C_{6}(T)^{\sigma_{m}},\quad\sigma_{m}=m+\lambda m(m-1)$, and $\lambda\approx 0.3$ in 3D case. We observed the same power law with $\lambda\approx 0.3$ for the single domain case, however, our results for the several domain case do not follow this power law and require further analysis.}.

The shape of the diffuse radial peak $\lvert I_{0}(q)\rvert=\langle \langle I(q,\varphi)\rangle_{\varphi}\rangle_{M}$ determines positional order correlation length, and
is described by a square-root Lorentzian (SRL) function \cite{Aeppli, Brock1}. Here, in addition to $\lvert I_{0}(q)\rvert$ we also fitted
the higher-order Fourier components $\lvert I_{n}(q)\rvert$ as a function of $q$ and shown in Fig.~\ref{Fig:XCCA1}(c) by a SRL function,
$\lvert I_{n}(q)\rvert=B+S\cdot\{\gamma_{n}/[(q-q_{0})^2+\gamma_{n}^2]\}^{1/2}$,
where $B$ is a background correction, $S$ is a scaling coefficient, $\gamma_{n}$ is half-width at half-maximum (HWHM) of the corresponding Lorentzian function,
and $q_{0}$ specifies the center of the function.
The results of SRL fits demonstrate good agreement with the experimental data.
For each Fourier component $\lvert I_{n}(q)\rvert$ we define positional correlation length as $\xi_{n}=2\pi/(\sqrt{3}\gamma_{n})$, where $\sqrt{3}\gamma_{n}$ is a HWHM of the SRL function.
For $n=0$, $\xi_{0}$ defines the conventional in-plane positional correlation length \cite{Pindak, Brock, Wim}, while for $n>0$ the correlation lengths $\xi_{n}$ characterize the short-range positional correlations corresponding to higher harmonics of BO order.

We determined the temperature evolution of positional correlation lengths $\xi_{n}$ for different $n$ from the corresponding SRL fits.
The resulting temperature dependence of $\xi_{0}$ obtained for the single domain case [see Fig.~\ref{Fig:XCCA2}(b)]
is in agreement with the previous results \cite{Pindak, Chou, Brock, Davey, Jeu}.
The correlation lengths $\xi_{n}$ associated with the higher harmonics of BO order
[see Fig.~\ref{Fig:XCCA2}(c)]
for each $n$ slightly decreases as the temperature grows, and the general character of decay is similar to that of $\xi_{0}$.
Remarkably, the correlation lengths $\xi_{n}$ have larger values than $\xi_{0}$, and consistently increase with a harmonic order $n$.
This observation indicates that in hexatics the positional order is getting more extended in space if one takes the higher order harmonics of BO order into consideration.
Such property is characteristic of well developed BO order, where the contribution of the higher order harmonics is larger.
In this situation coupling between the positional correlations and the BO order can not be neglected.
The evolution of the diffuse radial peak $\lvert I_{0}(q)\rvert$ in the vicinity of the hexatic-smectic transition in the presence of such a coupling has been studied by Aeppli and Bruinsma \cite{Aeppli} in the framework of the phenomenological XY model. It was shown that upon decreasing temperature the width of the diffuse radial peak diminishes simultaneously with the development of the BO order. However, no specific  dependence of the positional correlations on the BO order harmonics has been revealed. This question requires further theoretical and experimental analysis.

\section{Conclusions}

In summary, we applied XCCA to determine the BO order parameters in hexatic phase close to the hexatic-smectic phase transition.
Fourier analysis of the two-point CCFs provided a direct access to the parameters of BO and positional order in hexatic phase.
The BO parameters of the orders $n=6,12,18,24$ and $30$ were determined.
From the analysis of radial intensity profiles we determined the temperature dependence of the in-plane positional correlation length in hexatic membranes, that is in agreement with previous observations.
Our results show, that the correlation lengths $\xi_{n}$ for $n>0$, associated with higher harmonics of BO order in hexatic phase,
decrease in the vicinity of the hexatic-smectic phase transition similar to the in-plane positional correlation length $\xi_{0}$.
According to our results, the correlation lengths $\xi_{n}$
have larger values for the higher order Fourier components that is related to the interplay between positional and BO order.
Our results demonstrate the ability of XCCA to directly measure the parameters of BO order in smectic membranes.
This makes it a relevant tool for studying angular correlations and other order parameters in
various partially ordered molecular ensembles.
We expect that our results will stimulate further theoretical studies of ordering phenomena in hexatics and other LCs, and development of experimental techniques
that enable quantitative investigation of ordering in molecular films composed of multiple domains.

\section{Acknowledgments}

We thank Ewa Gorecka, C.C. Huang and Wim de Jeu for providing us hexatic materials and for stimulating discussions, M. Altarelli and J. Brock for fruitful discussions, and S. Funari for a careful reading of the paper.
We are grateful to Sergey Sulyanov and Pavel Dorovatovskii for preliminary characterization of hexatic liquid crystals at Kurchatov Synchrotron Centre, Moscow.
Part of this work was supported by BMBF Proposal 05K10CHG `'Coherent Diffraction Imaging and Scattering of Ultrashort Coherent Pulses with Matter`'
in the framework of the German-Russian collaboration `'Development and Use of Accelerator-Based Photon Sources`' and the Virtual Institute VH-VI-403 of the
Helmholtz Association.


\bibliography{XCCA_LC}


\end{document}